\documentclass[a4paper,12pt]{article}
\usepackage[margin=2cm, bottom = 2cm]{geometry}
\usepackage{multicol}
\usepackage{amsmath}
\usepackage{graphicx}
\usepackage{caption}

\usepackage[utf8]{inputenc}
\usepackage[english,russian]{babel}

\title{DEFINITION OF CYBERNETICAL NEUROSCIENCE
\author{Alexander Fradkov\\
Institute for Problems of Mechanical Engineering of RAS, \\
61, Bolshoy pr .V.O., Saint Petersburg 199178 Russia \\
St Petersburg University \\ 28 Universitetskii prospect, Peterhof,
St Petersburg 195904, Russia\\
E-mail: fradkov@mail.ru
}
}
\date{Sept. 07, 2024}

\begin{document}
\maketitle



A new scientific field is introduced and discussed, named cybernetical neuroscience, which studies mathematical models adopted in computational neuroscience by methods of cybernetics - the science of control and communication in a living organism, machine and society. It   also considers the practical application of the results obtained when  studying  mathematical models. The main tasks and methods, as well as some results of cybernetic neuroscience are considered.



\section{Introduction}
Over the past 20-25 years, the term computational neuroscience has become popular. It is used to describe a field of science at the intersection of neuroscience and a conglomerate of exact and technical sciences, such as mathematics, physics, computer science, programming and artificial intelligence, engineering sciences for the study of processes in the brain and nervous system of humans and animals. The term was introduced by Boston University professor Eric Schwartz 
who organized a multidisciplinary conference on this topic in 1985,
the proceedings of which were published in 1990 under the title Computational Neuroscience
\cite{Schwartz90}.

Since then, computational neuroscience has grown into a huge field and a number of directions have emerged within it, each with its own tasks and methods, see \cite{Feng03,Bower13,Trappenberg22}.
A number of methods are based on the use of mathematical models of neurobiological processes. Other methods do not rely explicitly on mathematical models, but some relationships from applied mathematics, physics, etc. are used in them.

Mathematical models of neuronal processes existed long before the formation of computational neuroscience as a science.
The most famous of them are based on the famous Hodgkin-Huxley model of 1952,
for which its authors received the Nobel Prize \cite{Hodgkin52}.
These are simplifications of the Hodgkin-Huxley model: the Morris-Lekar, Izhikevich, FitzHugh-Nagumo, Hindmarsh-Rose, and other models.
There are models derived from general principles, such as the "integrate-and-fire" model. There are models that come from related sciences, such as the Landau-Stewart model.

The above models describe the behavior of individual neurons or groups of neurons that oscillate synchronously between active and inactive states.
Each group can correspond, for example, to a certain region of closely located neurons,
whose states change in time in a coordinated manner.
If it is necessary to describe the movement of large ensembles of neurons, then network models are constructed in the form of a network of interconnected nodes with various interaction (connection) functions: linear (diffusion), nonlinear, with delays, etc.

Finally, there are many works studying the so-called neural mass models, proposed in 1995 \cite{Jansen95}, and describing neuronal populations. These models can also describe subpopulations, then to describe a large population, neural mass models are combined into networks \cite{Friston08}.
The listed theoretical models are compared with the results of experimental studies on mapping the interaction of neurons in the human brain (connectome) \cite{Tzourio02,connectome} and
represent the first steps towards solving a global problem: constructing mathematical models of the whole brain 

In recent years, research based on the application of cybernetics methods - the science of control and communication
in a living organism and a machine, as defined by its creator, the American mathematician Norbert Wiener in 1948 \cite{Wiener48}, has become increasingly popular in neurobiology. Cybernetics methods are based on the construction and use of mathematical models and the construction of direct and feedback links between the subject of the study (computer system) and the object of the study (in this case, the human or animal brain).
Cybernetics also uses mathematical models. However, they differ in that they contain input variables that set external influences on the brain and output variables that represent measurable quantities in the control object.

In the course of research, a new scientific direction in computational neuroscience is being formed,
which can be called ``cybernetic neurobiology''.
The purpose of this work is to outline the contours of this new scientific direction
Below are listed some of the main tasks, methods and results of cybernetical  neuroscience.

\section{Tasks of cybernetical  neuroscience}

Cybernetical neuroscience is aimed at solving the following main tasks:
1.  Analysis of the conditions for the models of neural ensembles to possess some special brain-like behaviors,
such as synchronization, desynchronization, spiking, bursting, solitons, chaos, chimeras and others.

2. Synthesis of external (control) inputs that create the special behaviors in the brain models.

3. Estimation of the state and parameters of the brain models based on the results of measuring input and output variables.

4. Classification of brain states and human intentions (using adaptation and machine learning methods).

5. Design of control algorithms that provide specified properties of a closed loop system consisting of an interacting controlled neural system and a controlling device.

 Recall that the controlled system in neurobiological research is the human nervous system or brain, and the control device can be implemented in some computer device.
For the entire system to work, the nervous system or brain must be connected to external computer communication devices called neurointerfaces (brain-computer interfaces).
However, in theoretical research, the cybernetic control system, i.e. both the human brain and the control device (algorithm), may not be physically implemented, but presented as a set of mathematical models.
In this case, solving the control synthesis problems for the model precedes the implementation of the control system for a real organism or brain, which provides neurobiologists with many new opportunities.

We also note that problem 1 is related to the problems of computational neurobiology,
differing from them in that in cybernetic research, input and output variables are identified in the models, through which the nervous system or brain interacts with the outside world. The sought conditions for the existence of the studied modes must be expressed in terms of input and output variables to ensure the possibility of their verification.

\section{Models of cybernetical  neuroscience}

As it was stated, cybernetic models must include input and output variables.
In the case of models described by vector differential state equations  of general form,
the cybernetic version of the mathematical  model has the form
 \begin{equation}
  \begin{cases}
    dx/dt = F(x,w,t),\\ y=h(x,w,t),
   \end{cases} \label{F}
\end{equation}
where $x=x(t)$ is $n$-dimensional vector of system state variables at time $t\ge 0$,
$w=w(t)$ is an $m$-dimensional vector of input (controling) variables,
$y=y(t)$ is an $l$-dimensional vector of output (measured) variables.
Vector functions $F(x,w,t),~h(x,w,t)$ satisfy conditions that ensure the existence of solutions to the system (\ref{F}) on the time interval of interest.

Below, we consider cybernetic versions of some mathematical models of
computational neuroscience. One of the most popular mathematical models of a neuron is the FitzHugh-Nagumo model, described by a system of two differential equations
\begin{equation}
  \begin{cases}
    \dot u = u - \frac{u^3}{3} - v + I_{ext},\\
    \dot v = \varepsilon(u - a - bv),
\end{cases} \label{FHN}
\end{equation}
where $u(t)$ is the neuron membrane potential (activator), $v(t)$ represents the cumulative effect of all slow ion currents responsible for restoring the resting potential of the nerve cell membrane (inhibitor). Parameters $a$ and $b$ determine the characteristics of the ion channel conductivity, $\varepsilon > 0$ is the relative change in the velocity of slow ion currents, $I_{ext}$ is the external current, the dot denotes differentiation with respect to time.

We will assume that the neuron membrane potential $u(t)$ is measurable. This means that $u(t)$ plays the role of an output variable. The model (\ref{FHN}) can be written in the form (\ref{F}), if we introduce the state vector $x={u,v}$, and the output variable $y=u$. The input of the model can be introduced in different ways. It can be the external current $I_{ext}$, if this variable is subject to purposeful change. The input can also be one of the model parameters $a$, $b$, or their function if this function can be purposefully changed. For simplicity, we will further assume that the input to the system is the external current $I_{ext}$.
\newline

If there is a need to study an ensemble of interacting neurons each of which is described by the FitzHugh-Nagumo model, then we can consider the networked model
 \begin{equation}
  \begin{cases}
    \dot u_i = u_i - \frac{u_i^3}{3} - v_i + I_{i,ext}+C_i\sum^{N}_{j=1}G_{ij}\varphi(u_i(t)-u_j(t-\tau_i)),\\
    \dot v_i = \varepsilon(u_i - a_i - b_iv_i), ~i=1,\ldots,N,
\end{cases} \label{FHN-N}
\end{equation}
where $u_i(t), v_i(t)$ are the variables of the model of the $i-$th network node, $C_i,a_i, \tau_i$ are the network parameters, $G$ is the matrix of information connections in the network. The model (\ref{FHN-N}) takes into account the delay in the transmission of information between neurons. Interest in network models increased after an atlas of the anatomical links of neurons in the human brain was developed in studies of the human connectome in the form of a network consisting of 90 nodes corresponding to important regions of the brain \cite{Tzourio02,connectome}.

The processes at each node can be described by various models, including the FitzHugh-Nagumo models.
In particular, in the work \cite{Gerster20} it is shown that  model (\ref{FHN-N}) can exhibit the processes similar to the development of epilepsy.

A more complex model that can also be obtained from the Hodgkin-Huxley model, is the Hindmarsh-Rose model, described by three differential equations:
\begin{equation} \label{HR}
\begin{cases}
    \dot p = q - ap^3 + bp^2 - n + I_{ext},\\
    \dot q = c - dp^2 - q,\\
    \dot n = r[s(p + p_0) - n],
\end{cases}
\end{equation}
where $p(t)$ denotes the neuron's membrane potential, $q(t)$ and $n(t)$ model the operation of the sodium-potassium pump,

The measured output of the model is the membrane potential $p(t)$, the input is the external current $I_{ext}$.

The models (\ref{HR}) can also be linked into networks that model the behavior of neural ensembles.


Relatively recently, the Landau-Stuart model has come into use for modeling the dynamics of neurons. It represents the normal form of systems undergoing a supercritical Andronov-Hopf bifurcation.
\begin{equation} \label{LS}
\frac{dZ}{dt} = Z[a + i\omega - |Z^2|]
\end{equation} where $Z(t)$ is a complex variable describing the state of the oscillator at time $t$.
The equation (\ref{LS}) was published by L.D. Landau in 1944 in the work \cite{Landau44} dedicated to the nature of the turbulence, and its derivation on the basis of hydromechanics was made by Stuart et al. in 1958, see \cite{Stuart58}.

Network versions of the Landau-Stuart model are used in the study and control of biological networks. In \cite{Hauptmann07} a method was proposed for controlling spatiotemporal patterns of synchronism in the networks of coupled Landau-Stuart oscillators in the form of delayed feedback. The Landau-Stuart network model in the version \cite{Hauptmann07}
has the form
\begin{equation} \label{LS-N}
    \frac{dZ_n}{dt} = Z_n[a + i\omega - |Z_n^2|] + K\sum_{p \neq n}^NC_{np}[Z_p(t - \tau_{np}) - Z_n(t)] + \beta\eta_1 + i\beta\eta_2, \forall n \in N,
\end{equation}
where $Z_n(t)$ is a complex variable describing the state of the $n$th node of the oscillator network at time $t$,
$\omega$ is the angular frequency of the oscillators' natural oscillations; $K$ is the strength of all network interactions; $C_{np}$ is the connectivity strength, $\tau_{np}$ are the conduction delays between each pair of brain areas $n$ and $p$; $\beta\eta_1 + i\beta\eta_2$ represents the added uncorrelated noise.

Control (stimulation) is applied using a small number of electrodes and induces spatiotemporal images even at low stimulation intensity.

\section{Methods of Cybernetical Neuroscience}

Since the goal of cybernetic neurobiology is to apply cybernetics methods to the study of neurobiological systems, its methods differ little from the methods of cybernetics, i.e. from the methods of control theory. In essence, the methodology is the same as for cybernetic physics \cite{F05UFN,F07}. For the synthesis of control algorithms,
nonlinear control methods (gradient method and speed gradient method),
adaptive and optimal control methods, pattern recognition and machine learning methods are used.

\section{Results of cybernetical neuroscience: model study}

\subsection{Regulation, tracking, synchronization, chaos control}

The results that relate to the field of cybernetic neurobiology belong to two classes: results concerning the control of processes in models of neurons and neural networks
and results concerning control based on real data. Let us first list the results on the control of models of neurons and neural networks.

These results have theoretical significance, demonstrating the fundamental possibility of controlling neuronal processes under the assumption that the models describe real processes well enough, the output variables are measurable, and the input variables are modifiable.

The control problem, in addition to the model of the control object, also includes a description of the control goal. The goal may be the requirement, common in cybernetics, of approaching the process to a given state or a given trajectory (regulation or tracking). The tasks of synchronization of processes in different parts of systems and the occurrence of oscillations, their chaotization, etc. can also be set.

The first results on the control of neuron and neural network models were obtained in the 1990s and related to the control of chaos and synchronization. In the work \cite{Carroll95}, an algorithm for pulse control of the synchronization of two FHN models was proposed based on modeling and analogy between neuronal and electrical processes.
In the work \cite{Dragoi98}, an algorithm for controlling a chain of FHN neurons was proposed with the aim of bringing together (synchronizing) the oscillations of each neuron with the oscillations of the ``reference'' neuron. The stability of the synchronization process was established in a certain region of initial conditions based on a linear approximation.

The authors of the work \cite{Plotnikov16a} proposed algorithms for synchronizing a heterogeneous network of diffusion-coupled models of FHN neurons with a hierarchical architecture, based on the speed gradient method.
Synchronization conditions were obtained based on the Lyapunov function method.
Similar results were obtained for adaptive control algorithms that do not require precise knowledge of the parameters of neuron models \cite{Plotnikov16b}, for networks of arbitrary structure
\cite{Plotnikov19a} and networks with delays in connections \cite{Plotnikov19b},
as well as for the desynchronization problem \cite{Plotnikov19c},
important for the treatment of a number of mental illnesses such as Parkinson's disease, tremor, etc.

The Lyapunov function method and the speed gradient method were also successfully applied
to the construction and study of control algorithms in synchronization problems and chaos control in
Hindmarsh-Rose models and their networks \cite{Plotnikov21,Semenov22}, in problems of
oscillation control in neuromass models, as well as in problems of adaptive control of Landau-Stewart oscillator networks \cite{Selivanov12,Selivanov14}.

\subsection{Estimation of the state and parameters of models}

The problem of estimating the state and parameters of a neural ensemble model is important for ensemble control based on measurable data. In addition, knowledge of the parameters and state of the network is important for better judgment of its behavior and properties. There are many works on estimating the parameters of a single neuron model, and most of them use stochastic approaches \cite{Jensen12,Che12,Doruk19}
There are works \cite{Dong15,Rudi22}, where artificial neural networks are used to estimate the parameters of FHN models.
In \cite{Rybalko23}, an algorithm for estimating the state and parameters of a pair of FHN neurons based on the speed gradient method and filtering is proposed and justified.
The speed gradient method was also applied to estimating the parameters of the Hindmarsh-Rose neuron model \cite{Kovalchukov22,Kovalchukov22a} and the neural mass model \cite{Plotnikov24}.
Other approaches to estimating the parameters of neuron models are presented in \cite{Postoyan12,Zhao16,Dong19,Wang19}.

\section{Results of Cybernetical Neuroscience: Study of Real Data}

\subsection{Classification of brain states and diagnostics}

 Here, pattern recognition methods and machine learning methods are used, which are often attributed to the field of artificial intelligence. The classification problem is as follows.
The results of measuring the state of a finite set of $N$ neurobiological objects are given, each of which belongs to one of $M$ classes. It is required to construct a set of decision rules that allow one to determine, based on the measured data, which class a new, classified object belongs to.
Such problems are typical for medical diagnostics, where cybernetic pattern recognition methods
have been used for a long time. In neurophysiology and psychiatry, recognition and machine learning methods are actively used, see \cite{Mueller10, Lebedev14, Boyko22, Zubrikhina22, Yoon22, Shanarova23}.
Both well-known statistical methods (discriminant analysis, principal component analysis (PCA), independent component analysis (ICA), random forests) and deterministic machine learning methods are used.
Works are appearing on the application of new approaches to neuroscience, for example, the method of target inequalities \cite{Lipkovich22}.

\subsection{Neurofeedback control}

Neurofeedback (NFB) (or biological feedback (BFB)) is the most effective approach to the interaction of the human brain with an external control device and the most promising cybernetic approach in neurobiology and neurophysiology.
 During a neurofeedback experiment, the subject is presented with information about the state and desired change of certain physiological parameters. The basic principle of cybernetics - feedback (display of information about the results of activity), serves as a "mirror" in which one can see physiological parameters otherwise inaccessible to consciousness and regulate the parameters of electrical activity of the brain. The implementation of neurofeedback requires the presence of a neurointerface - a device that provides information exchange between the brain and the computer in real time. Typically, a noninvasive neurointerface uses electroencephalography (EEG) data reflecting changes in the electric field potential on the subject's scalp. Some of the current EEG parameters (or their combination) \cite{Kropotov10} are presented to the subject in the form of, for example, a visual stimulus (the height of a column on the screen, the brightness of the screen) with the task of changing these parameters in the desired direction. In such a paradigm, the subject, focusing on the NOS signal, tries to remember the relationship between the parameter and his state. The EEG parameters and the localization of the electrodes that form the NOS protocol are selected depending on the task \cite{Kamiya68}.

The task of generating a neurofeedback signal is very complex, since at the moment there are no clear rules for presenting a stimulus that must be followed in order to help the subject cope with the task most effectively (for example, in terms of time spent)\cite{Holten09,Kropotov10}.
One of the promising ways of developing neurofeedback is the use of adaptive mathematical models of brain activity and the mathematical apparatus of the theory of adaptive control proposed in \cite{Ovod12, Plotnikov19} \cite{MNF00, FFYa81}.

\section{Conclusion}

The application of cybernetics and control theory methods to problems of neurobiology and neurophysiology has great potential and the number of publications in this area is growing rapidly.
General publications and reviews appear \cite{Wilson22, Howlett24}.
This paper is the first attempt to structure and systematize this area, to
outline the contours of its methodology.


\begin{thebibliography}{99}
\bibitem{Schwartz90}
  Computational neuroscience. Ed.  Eric Schwartz. Cambridge, Mass: MIT Press, 1990.

 \bibitem{Feng03}
   Computational Neuroscience: A comprehensive approach. Ed. Jianfeng Feng.
CRC Press,  2003.

 \bibitem{Bower13}
  20 years of Computational neuroscience. Ed. James M. Bower. Berlin: Springer, 2013.

\bibitem{Trappenberg22}
 Trappenberg, Thomas P.
 Fundamentals of Computational Neuroscience.  3rd ed. United States: Oxford University Press Inc, 2022


\bibitem{Hodgkin52}
Hodgkin, A.L., Huxley, A.F.: A quantitative description of membrane current and
its application to conduction and excitation in nerve. J. Physiol. 117, 500–544
(1952)

 \bibitem{Jansen95}
Jansen, B., \& Rit, V. (1995). Electroencephalogram and visual evoked potential
generation in a mathematical model of coupled cortical columns. Biological
Cybernetics, 73, 357–366.


 \bibitem{Friston08}
C.C. Chen, S.J. Kiebel, and K.J. Friston
Dynamic causal modelling of induced responses.
NeuroImage 41 (2008), 1293–1312.

\bibitem{Wiener48}
Wiener~N. Cybernetics: or, control and communication in the animal and
  the machine.  MIT Press, 1948.

  \bibitem{Tzourio02}
  N. Tzourio-Mazoyer, B. Landeau, D. Papathanassiou, F. Crivello, O. Etard,
N. Delcroix, B. Mazoyer, and M. Joliot,
Automated anatomical labeling of activations in SPM using a macroscopic anatomical
 parcellation of the MNI MRI single-subject brain.
 Neuroimage 15, 273 (2002).

   \bibitem{connectome}
 Human Connectome Projects
 https://www.humanconnectome.org/

   \bibitem{Gerster20}
 M. Berster, R. Berner, J. Sawicki, A. Zakharova, A. Škoch, J. Hlinka,
K. Lehnertz, and E. Schöll, “FitzHugh–Nagumo oscillators on complex
networks mimic epileptic-seizure-related synchronization phenomena,”
Chaos: An Interdisciplinary Journal of Nonlinear Science 30, 123130
(2020).


\bibitem{Landau44}
Landau L. On the problem of a turbulence. Dokl. Akad. Nauk SSSR Vol. 44, No. 8, pp. 339-349, 1944.

\bibitem{Stuart58}
Stuart, J. T. On the non-linear mechanics of hydrodynamic stability.
Journal of Fluid Mechanics, 4(1), 1-21, 1958.

\bibitem{Hauptmann07}
C. Hauptmann, O. Omel‘chenko, O. V. Popovych, Y., Maistrenko, and P. A. Tass,
Control of spatially patterned synchrony with multisite delayed feedback.
Phys. Rev. E 76, 066209 (2007).

\bibitem{Freyer12}
Freyer F, Roberts J.A, Ritter P, Breakspear M (2012) A Canonical Model of Multistability and Scale-Invariance in Biological Systems.
PLoS Comput Biol 8(8): e1002634.

\bibitem{F05UFN}
A.L. Fradkov. Application of cybernetic methods in physics.
       Physics-Uspekhi, Vol. 48 (2), 2005, 103-127)

\bibitem{F07}
Fradkov A.L.Cybernetical physics: from control of chaos to quantum control.
  Berlin Heidelberg: Springer-Verlag, 2007.
  
\bibitem{Carroll95}
Carroll T.
Synchronization and complex dynamics in pulse-coupled circuit models of neurons
Biol. Cybern. 73, 553-559, 1995.

\bibitem{Dragoi98}
Dragoi, V., Grosu, I. Synchronization of Locally Coupled Neural Oscillators.
Neural Processing Letters 7, 199–210, 1998.

\bibitem{Plotnikov16a}
S. A. Plotnikov, J. Lehnert, A. L. Fradkov, E. Schoeell.
Synchronization in heterogeneous FitzHugh-Nagumo networks with hierarchical architecture.
Phys. Rev. E 94, 012203 (2016).

\bibitem{Plotnikov16b}
Plotnikov S.A., Lehnert J., Fradkov A.L., Schoell E.
Adaptive control of synchronization in delay-coupled heterogeneous networks
of FitzHugh-Nagumo nodes. International Journal of Bifurcation and Chaos,
Vol. 26, No. 4 (2016) 1650058 (14 pages).

\bibitem{Plotnikov19a}
 Sergei A.Plotnikov, Alexander L.Fradkov
On synchronization in heterogeneous FitzHugh–Nagumo networks
Chaos,Solitons and Fractals, 121(2019) 85–91.

\bibitem{Plotnikov19b}
Plotnikov, Sergei, Fradkov, Alexander L
On Synchronization in FitzHugh-Nagumo Networks with Small Delays.
Proc. 2018 Europ. Contr. Conf. (ECC 2018), Limassol, 13-15 June, 2018, pp.2052-2056.

\bibitem{Plotnikov19c}
Plotnikov, S.A., Fradkov A.L.
Desynchronization Control of FitzHugh-Nagumo Networks with Random Topology.
IFAC-PapersOnLine, Volume 52, Issue 16, 2019, Pages 640-645.

 \bibitem{Plotnikov21}
Plotnikov, S. Synchronization conditions in networks of Hindmarsh–Rose systems.
Cybern. Phys. 2021, 10, 254–259.

\bibitem{Semenov22}
 Semenov D.M., Plotnikov S.A., Fradkov A.L.
Controlled synchronization in regular delay-coupled networks of Hindmarsh-Rose neurons.
2022 6th Scientific School Dynamics of Complex Networks and their Applications (DCNA).
IEEE,  Kaliningrad, 2022.

\bibitem{Selivanov12}
Selivanov A.A., Lehnert J, Dahms T., Hovel P., Fradkov A.L., Schoell E.
Adaptive synchronization in delay-coupled networks of Stuart-Landau oscillators.
Phys.Rev. E 85, 016201 (2012).

\bibitem{Selivanov14}
J. Lehnert, P. Hovel, A. A. Selivanov, A. L. Fradkov, and E. Schoell.
Controlling cluster synchronization by adapting the topology,
 Physical Review E 90, 042914 (2014).

\bibitem{Jensen12}
A. C. Jensen, S. Ditlevsen, M. Kessler, and O. Papaspiliopoulos, “Markov
chain Monte Carlo approach to parameter estimation in the FitzHugh-
Nagumo model, Physical Review E 86, 041114 (2012).

\bibitem{Che12}
Y. Che, L. Geng, C. Han, S. Cui, and J. Wang, Parameter estimation
of the FitzHugh-Nagumo model using noisy measurements for membrane
potential, Chaos: An Interdisciplinary Journal of Nonlinear Science 22,
023139 (2012).

 \bibitem{Doruk19}
R. O. Doruk and L. Aboshar, Estimating the parameters of
FitzHugh–Nagumo neurons from neural spiking data, Brain Sci. 9,
364 (2019).

 \bibitem{Dong15}
X. Dong and C. Wang, Identification of the FitzHugh–Nagumo model dynamics
via deterministic learning, International Journal of Bifurcation and
Chaos 25, 1550159 (2015).

 \bibitem{Rudi22}
J. Rudi, J. Bessac, and A. Lenzi, Parameter estimation with dense and
convolutional neural networks applied to the FitzHugh-Nagumo ODE, in
Mathematical and Scientific Machine Learning (PMLR, 2022) pp. 781–808.

 \bibitem{Rybalko23}
Rybalko A., Fradkov A.
Identification of Two-Neuron FitzHugh-Nagumo Model Based on the Speed-Gradient and Filtering
Chaos 2023, Vol.33, Is, 8. 083126.

 \bibitem{Kovalchukov22}
Fradkov A.L., Kovalchukov A. and B. Andrievsky
 Parameter Estimation for Hindmarsh–Rose Neurons
Electronics 2022, 11(6), 885;

 \bibitem{Kovalchukov22a}
 Kovalchukov A.A., A.L. Fradkov
Speed-Gradient approach to Hindmarsh-Rose model identification based on membrane potential measurements
2022 6th Scientific School Dynamics of Complex Networks and their Applications (DCNA). IEEE,
Kaliningrad, Russian Federation, 2022..

 \bibitem{Plotnikov24}
Plotnikov, S.A. Fradkov, A.L.
Adaptive Parameter Identification for a Class of Neural Mass Models with Application
 to Ergatic Systems. Mekhatronika, Avtomatizatsiya, Upravlenie, 2024, 25(1), pp.13–18.


 \bibitem{Postoyan12}
Postoyan, R.; Chong, M.; Neši´c, D.; Kuhlmann, L. Parameter and state estimation for a class of neural mass models. In
Proc. 51st IEEE Conf. Decision Control (CDC 2012), Maui, HI, USA, 10–13 December 2012; pp. 2322–2327.

 \bibitem{Zhao16}
  Zhao, J.; Aziz-Alaoui, M.A.; Bertelle, C.; Corson, N. Sinusoidal disturbance induced topology identification of Hindmarsh–Rose neural networks. Sci. China Inf. Sci. 2016, 59, 112205.

 \bibitem{Dong19}
 Dong, X.; Si,W.;Wang, C. Global Identification of FitzHugh–Nagumo Equation via Deterministic Learning and Interpolation. IEEE Access 2019, 7, 107334–107345.

\bibitem{Wang19}
  Wang, L.; Yang, G.; Yeung, L. Identification of Hindmarsh–Rose Neuron Networks Using GEO Metaheuristic. In Proceedings of the Second International Conference on Advances in Swarm Intelligence-Volume Part I, Chiang Mai, Thailand, 26–30 July 2019;
Springer: Berlin/Heidelberg, Germany, 2011; pp. 455–463.

\bibitem{Howlett24}
Jonathon R. Howlett, · Martin P. Paulus.
Out of control: computational dynamic control dysfunction in stress and anxiety-related disorders.
Discover Mental Health (2024) 4:5.

\bibitem{Mueller10}
 A. Mueller, G. Candrian, J. D. Kropotov, V. A. Ponomarev, and G. M. Baschera, Classification of ADHD patients on the basis of independent ERP components using a machine learning system, Nonlinear Biomed Phys, vol. 4 Suppl 1, p. S1, Jun 3 2010.

\bibitem{Lebedev14}
 Lebedev A., Westman E., Van Westen G., Kramberger M., Lundervold A., Aarsland D., Soininen H., Kłoszewska I., Mecocci P., Tsolaki M., et al.
Random Forest ensembles for detection and prediction of Alzheimer’s disease
with a good betweencohort robustness. NeuroImage Clin. 2014;6:115–125

\bibitem{Boyko22}
 Boyko Maria, Abramova Olga, Yarkin Vyacheslav,  Sharaev, Maxim et al.  Machine learning approaches to Mild Cognitive Impairment detection based on structural MRI data and morphometric features. Cognitive Systems Research. 78, (2022).

\bibitem{Zubrikhina22}
 Maria Zubrikhina, Dmitrii Masnyi,  Maxim Sharaev et al. Autoencoders with
deformable convolutions for latent representation of EEG spectrograms in classification tasks,
 Proc. SPIE 12701, 15th International Conference [on Machine Vision (ICMV 2022)

\bibitem{Yoon22}
 Yoon J., Kang C., Kim S., Han J. D-vlog: Multimodal Vlog Dataset for Depression Detection. Proc. Conf. AAAI Artif. Intell.
2022; 36:12226–12234

\bibitem{Shanarova23}
 Shanarova N., Pronina, M., Lipkovich, M. Müller, A.,Kropotov, J.
Application of Machine Learning to Diagnostics of Schizophrenia Patients Based on Event-Related Potentials,  Diagnostics, 2023, 13(3), 509.

\bibitem{Lipkovich22}
M. Lipkovich. Yakubovich's method of recursive objective inequalities in machine learning. IFAC-PapersOnLine, 2022, vol. 55, no. 12, pp. 138-143.

\bibitem{Ovod12}
 Овод И.В., Осадчий А.Е., Пупышев А.А., Фрадков А.Л. Формирование нейрообратной связи
   на основе адаптивной модели активности головного мозга. Нейрокомпьютеры, 2012, 2, С.36-41.

\bibitem{Plotnikov19}
 Plotnikov S.A., Semenov D.M., Lipkovich M., Fradkov A.L.
Artificial intelligence based neurofeedback.
Cybernetics And Physics, 2019, Vol. 8, Is. 4, 287–291,

\bibitem{Kropotov10}
  Кропотов Ю.Д. Количественная ЭЭГ, Донецк:  изд-во «Заславский» 2010.

\bibitem{Kamiya68}
 Kamiya J. Conscious control of  brain waves. Psychology Today, 1968,1, 56–60.

\bibitem{Holten09}
Holten V. Bio and neurofeedback applications in stress regulation. Neuroscience \& Cognition, track Behavioural Neuroscience, July 2009.

\bibitem{MNF00}
Мирошник И. В., Никифоров В.О., Фрадков А.Л., Нелинейное и адаптивное управление сложными динамическими системами, СПб. Наука, 2000.

\bibitem{FFYa81}
 Фомин В.Н., Фрадков А.Л., Якубович В.А., Адаптивное управление динамическими объектами. – М.: Наука, Физматлит, 1981.

\bibitem{Wilson22}
Wilson D., Moehlis J.
Recent advances in the analysis and control of large populations of neural
oscillators.
Annual Reviews in Control, Volume 54, , 2022,  327-351.

  \end{thebibliography}
\end{document}